# Silicon nanophotonic chips featuring Sagnac loop reflectors as a basis for advanced optical spectral filters

David J. Moss

Optical Sciences Centre, Swinburne University of Technology, Hawthorn, VIC 3122, Australia.

## ABSTRACT

We present and investigate theoretically photonic integrated filters based on 2 Sagnac coupled loop reflectors (SLRs) that are arranged in a self-coupled optical waveguide. We recently presented photonic integrated filters based on coupled and cascaded SLRs. In this paper, we advance this field by investigating a unique approach of employing coupled SLRs formed by self-coupled waveguides. This allows us to achieve high performance filter functions including Fano-like resonances and wavelength interleaving with a simpler design and a higher fabrication tolerance by tailoring coherent mode interference in the device. Our design takes into account the device fabrication issues as well as the requirements for practical applications. As a guide for practical device fabrication, an analysis of the impact of the structural parameters and fabrication tolerance on each filter function is also provided. The Fano-like resonances show a low insertion loss (IL) of 1.1 dB, a high extinction ratio of 30.2 dB, and a high slope rate (SR) of 747.64 dB/nm. The combination of low IL and high SR promises this device for Fano resonance applications. Our device also can achieve wavelength de-interleaving function with high fabrication tolerance which is advantageous for applications such as optical interleavers that require a symmetric flat-top filter shape. Optical de-interleavers and interleavers are key components for optical signal multiplexing and demultiplexing for wavelength division multiplexing optical communication systems. Versatile spectral responses with a simple design, compact device footprint, and high fabrication tolerance make this approach highly promising for flexible response shaping for many applications.

## 1. INTRODUCTION

Photonic integrated resonators have enabled a many of optical devices such as modulators, filters, switches, sensors, and logic gates, because of their small footprint, high scalability and flexibility [1-4]. Compared to the resonators based on photonic crystals [5] and gratings [6] that have sub-wavelength dimensions, devices based on directional-coupled wire waveguides with long cavity lengths yield much lower free spectral ranges (FSRs), matching the spectral spacings of current optical communication systems based on wavelength division multiplexing (WDM), thus making them widely applicable to these systems. In addition, sub-wavelength dimensions of photonic crystal cavity and Bragg grating structures are more prone to fabrication tolerances as compared with directional-coupled wire waveguides. Ring resonators (RRs), and Sagnac loop reflectors (SLRs), which are essential building blocks for IPRs, are made up of directional couplers. Unlike RRs, which only allow for unidirectional light propagation, SLRs allow for bidirectional light propagation as well as mutual coupling between light travelling in opposing directions, resulting in a more versatile coherent mode interference and spectral response. Furthermore, a standing-wave (SW) resonator made up of cascaded SLRs has a cavity length about half that of a traveling wave (TW) resonator made up of a ring resonator with the same FSR, allowing for a more compact device footprint.

We investigated integrated photonic filters based on cascaded SLRs [7, 8] and coupled SLRs [9, 10] in our previous work. Here, we advance this field by presenting the novel approach of using two coupled SLRs with a feedback loop formed by a self-coupled wire waveguide that yield different response shapes including Fano-like resonances and wavelength de-interleaving [11].

In our design, we take into account the device fabrication issues experienced in Refs. [7, 8] as well as the needs for practical applications. As a guide for practical device fabrication, an analysis of the influence of structural parameters and fabrication tolerance is also provided.

## 2. DEVICE CONFIGURATION

The proposed structure is illustrated schematically in Fig. 1, which consists of two inverse-coupled SLRs with a feedback loop formed by a single self-coupled wire waveguide. Table 1 details the device's structural parameters. To simplify the discussion, we assume that $L_{SLR1} = L_{SLR2} = L_{SLR}$. The spectral response of the device is calculated using the scattering matrix method [7, 9]. In the device model, we use waveguide group index of $n_g = 4.3350$ (transverse electric (TE) mode) and propagation loss of $\alpha = 55$ m$^{-1}$ (i.e., 2.4 dB/cm), which are in line with our previously fabricated silicon-on-insulator (SOI) devices [7, 8, 12]. The device is designed based on, but not limited to, the SOI platform.

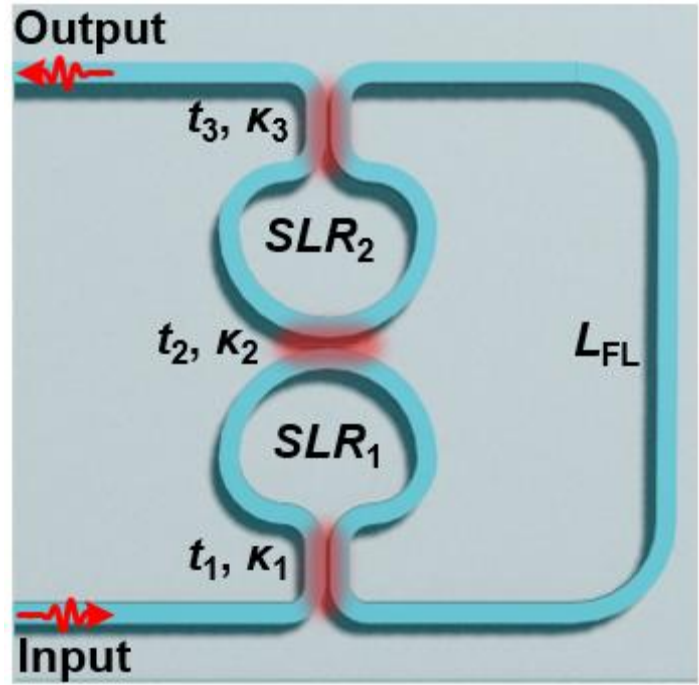

Figure 1. Schematic configuration of device. The definitions of $t_i$ ($i$ = 1, 2, 3), $L_{SLRi}$ ($i$ = 1, 2), and $L_{FL}$ are given in Table 1.

Table 1. Definitions of device structural parameters

| Waveguides | Length | Transmission factor [a] | Phase shift [b] |
|---|---|---|---|
| Feedback loop between SLRs ($i$ = 1, 2) | $L_{FL}$ | $a_f$ | $\varphi_f$ |
| Sagnac loop in $SLR_i$ ($i$ = 1, 2) | $L_{SLRi}$ | $a_{si}$ | $\varphi_{si}$ |
| Directional couplers | | Field transmission coefficient [c] | Field cross-coupling coefficient [c] |
| Coupler in $SLRs$ ($i$ = 1, 2) | | $t_i$ | $\kappa_i$ |
| Coupler between $SLR_s$ | | $t_2$ | $\kappa_2$ |

[a] $a_f = \exp(-\alpha L_{FL} / 2)$, $a_{si} = \exp(-\alpha L_{SLRi} / 2)$, $\alpha$ is the power propagation loss factor.
[b] $\varphi_f = 2\pi n_g L_{FL} / \lambda$, $\varphi_{si} = 2\pi n_g L_{SLRi} / \lambda$, $n_g$ is the group index and $\lambda$ is the wavelength.
[c] $t_i^2 + \kappa_i^2 = 1$ for lossless coupling are assumed for all the directional couplers

In the following sections, mode interference in the device is tailored to achieve high-performance filtering functions, including Fano-like resonances and wavelength de-interleaving.

## 3. FANO-LIKE RESONANCES

Fano resonances are a fundamental physical phenomenon demonstrating an asymmetric spectral lineshape arising from quantum interference between discrete and continuum states [13, 14]. They have enabled many applications such as sensing, topological optics, data storage, and optical switching, because of their unique physics that can yield extremely narrow spectral linewidths [13-15]. Here, the spectral response of the device in Fig. 1 is engineered to achieve Fano-type resonances with low insertion loss (IL) and high slope rates (SRs). The power transmission and reflection spectra are shown in Fig. 2(a). The device structural parameters are $L_{SLR}$ = 100 μm, $t_1 = t_3$ = 0.82, $t_2$ = 0.92, and $L_{FL}$ = 300 μm. Clearly, there are periodic Fano resonances with identical asymmetric resonant lineshapes in each period at the output port. The very high uniformity of the response shape of the resonator is ideal for WDM systems. A zoom-in view of Fig. 2(a) is shown in Fig. 2(b), together with another curve showing the corresponding result for another device with the same structural parameters except for a different $t_2$ = 1. As can be seen, when $t_2$ = 1, there is no Fano resonance, distinguishing between the device in Fig. 1 and the two cascaded SLRs in Ref. [16]. The Fano resonances in Fig. 2(a) show a high extinction ratio (ER) of 30.2 dB and a high SR (defined as the ratio of the ER to the wavelength difference between the resonance peak and notch) of 747.64 dB/nm.

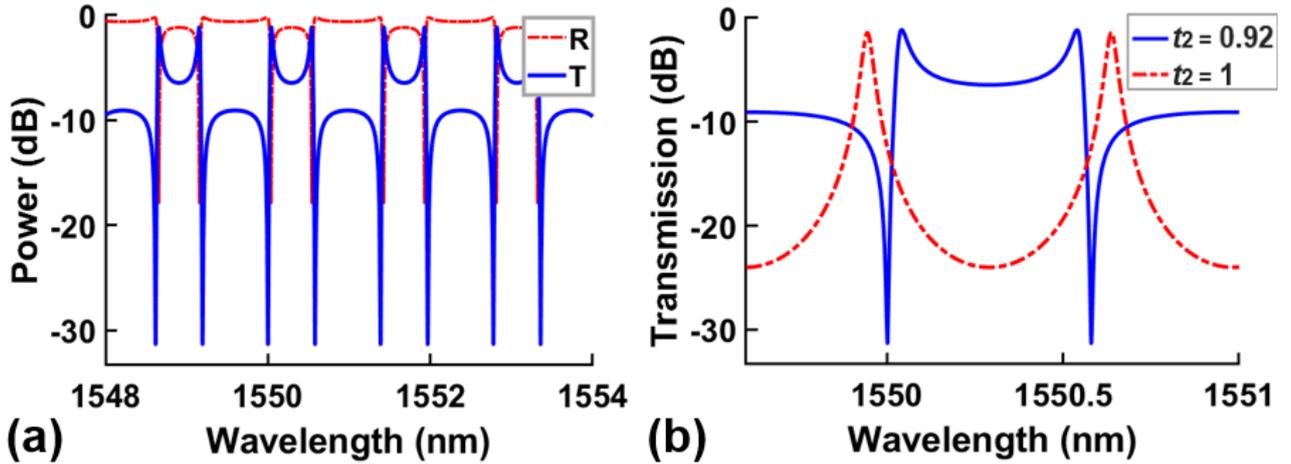

Figure 2. (a) Power transmission and reflection spectra when $L_{SLR}$ = 100 μm, $L_{FL}$ = 300 μm, $t_1 = t_3$ = 0.82, and $t_2$ = 0.92. T: Transmission spectrum at output port. R: reflection spectrum at input port. (b) Power transmission spectra at output port for $t_2$ = 0.92 and $t_2$ = 1. In (b), the structural parameters are kept the same as those in (a) except for $t_2$.

The Fano-like resonances have a performance enabled by the coupled SLRs that we presented in earlier work [9, 10] and the devices in Fig. 1 are listed in Table 2. For comparison, the device structural parameters ($L_{SLRi}$, $n_g$, and α) of all the three structures were kept the same except for the transmission coefficients ($t_i$) that were tuned to obtain the highest SR for each structure. As compared with previous devices, the device presented here has a much lower IL of 1.1 dB, along with a slightly improved SR. The combination of high SR and low IL promises this device for Fano resonance applications. We note that a low IL of 1.1 dB is outstanding among the reported Fano-resonance devices on the SOI platform [17, 18], which makes the device here more attractive for practical applications.

Table 2. Performance comparison of Fano-like resonances generated by different SLR-based devices.

| Device structure | IL (dB) | ER (dB) | SR (dB/nm) | FSR (GHz) | Ref. |
|---|---|---|---|---|---|
| Two parallel WC-SLRs [a] | 6.3 | 13.9 | 389 | 692.02 | [9] |
| Three zig-zag WC-SLRs [b] | 3.7 | 63.4 | 721.28 | 230.68 | [10] |
| Device in Fig. 1 | 1.1 | 30.2 | 747.64 | 173 | This work |

[a] WC-SLRs: waveguide coupled SLRs.
[b] For comparison, the length of the SLRs ($L_{SLRi}$, $i$ = 1–3) and the connecting waveguide ($L_i$, $i$ = 1–4) is slightly changed from 115 μm in [10] to 100 μm.

In Figs. 3(a)–(c), we investigate the impact of the parameters of the device including length variations of feedback loop ($\Delta L_{FL}$) and the $t_i$ ($i$ = 1–3) on the performance of the device. In each figure, we changed only one structural parameter, keeping the others the same as those in Fig. 2 (a). In Figs. 3(a)–(c), (i) shows power transmission spectra and (ii) shows the corresponding IL and SR for different $t_i$ ($i = 1–3$), and $\Delta L_{FL}$, respectively. The SR decreases with $t_i$ ($i = 1, 3$), whereas the IL first decreases with $t_i$ ($i = 1, 3$) after which it stays almost constant. The SR decreases with $t_2$, while the IL behaves oppositely, implying that both parameters can be enhanced by increasing the coupling strength between $SLR_1$ and $SLR_2$. The filter shape (Fig. 3(c)) does not change whereas the Fano-like resonance shifts to the red with increasing $\Delta L_{FL}$. This implies that the wavelength resonances could be tuned with thermo-optic micro-heaters [18] or carrier-injection electrodes [19] along a feedback loop to vary the phase shift.

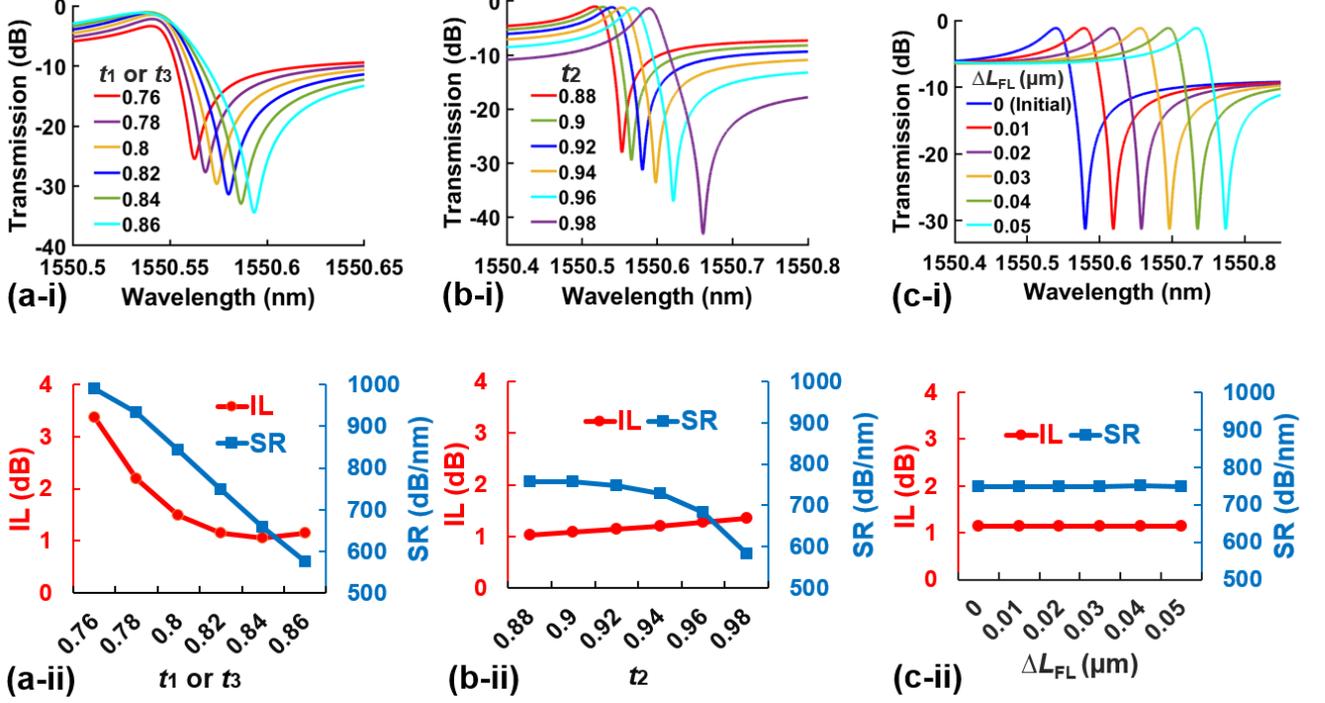

Figure 3. (a)–(c) Power transmission spectra (i) and the corresponding IL and SR (ii) for different $t_i$ ($i = 1–3$) and $\Delta L_{FL}$ respectively. In (a)–(c), the structural parameters are kept the same as those in Fig. 2(a) except for the varied parameters.

## 4. WAVELENGTH DE-INTERLEAVING FUNCTION

Optical de-interleavers and interleavers are key components for signal demultiplexing/multiplexing in wavelength division multiplexing telecommunication systems [20, 21]. Here, we optimize the device spectral response of Fig. 1 to demonstrate wavelength de-interleaving. Flat-top spectral response of de-interleavers minimize the filtering distortions and group delay variation and high ER minimize signal crosstalk between adjacent channels [22]. Fig. 4(a) shows the power transmission and reflection spectra when the device structural parameters are $L_{SLR}$ = 100 μm, $L_{FL}$ = 300 μm, $t_1$ = 0.992, and $t_2 = t_3$ = 0.95. The IL, ER, and 3-dB bandwidth for the passband at output port are 0.36 dB, 12.7 dB, and 83.65 GHz, respectively. The IL, ER, and 3-dB bandwidth for the reflection spectrum at input port are 0.33 dB, 12 dB, and 91.9 GHz, respectively. As compared with flat-top filters based on cascaded ring resonators [23], ring-assisted Mach-Zehnder interferometers [24], and cascaded SLRs [7], our device can achieve the same level of filtering flatness with fewer subunits.

We investigate the impact of varying $t_i$ ($i = 1–3$) in Figs. 4(b)–(d), showing only the spectral response of the output port for simplification. As $t_1$ increases in Fig. 4(b), the passband ER decreases while the flatness at the top improves, reflecting their trade-offs. In Figs. 4(c)–(d), the passband bandwidths increase with $t_2$, $t_3$ whereas the ER has the opposite behaviour.

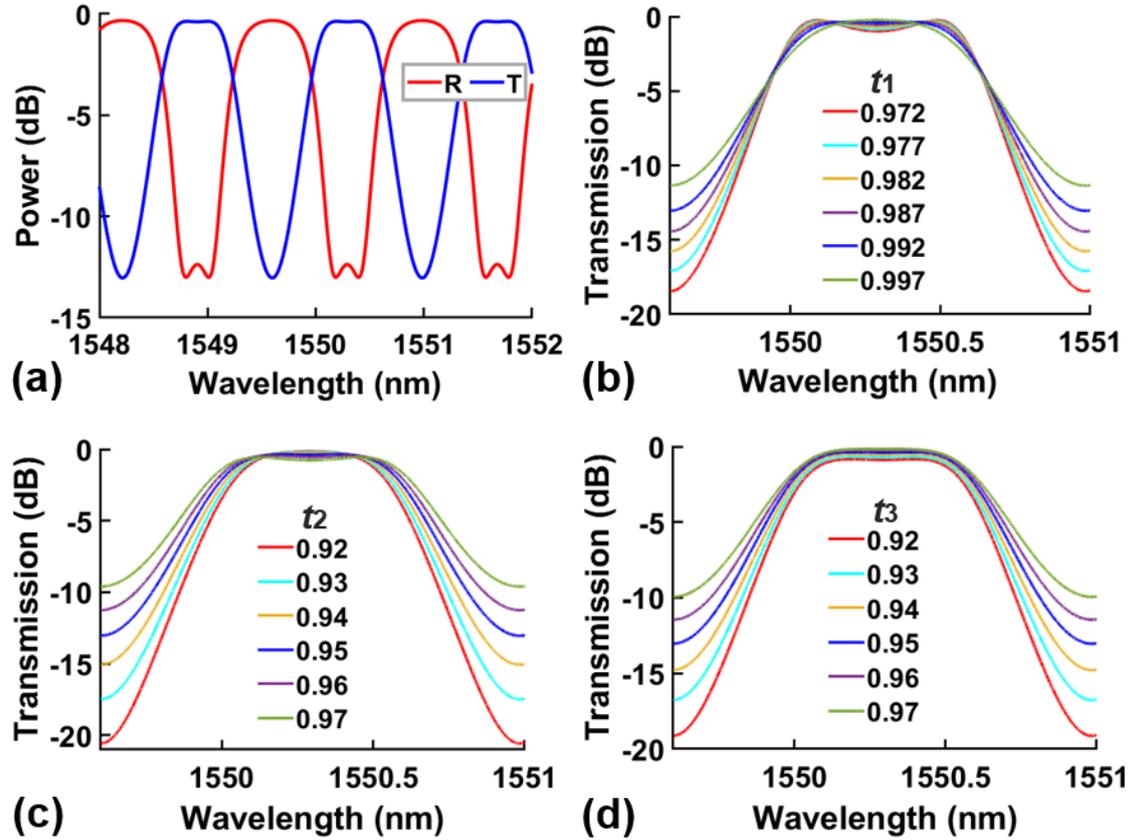

Figure 4. (a) Power transmission and reflection spectra of the device when $L_{SLR}$ = 100 μm, $L_{FL}$ = 300 μm, $t_1$ = 0.992, and $t_2 = t_3$ = 0.95. T: Transmission spectrum at output port. R: reflection spectrum at input port. (b)–(d) Power transmission spectra for different $t_i$ ($i$ = 1–3), respectively. In (b)–(d), the structural parameters are kept the same as those in (a) except for the varied parameters.

We also investigate the impact of varied $\Delta L_{SLRi}$ ($i$ = 1, 2) and $\Delta L_{FL}$ in Figs. 5(a)–(c), respectively. In Figs. 5(a)–(c), as $\Delta L_{SLRi}$ ($i$ = 1, 2) or $\Delta L_{FL}$ increases, the filter shape remains unchanged while the resonance redshifts. Since the resonant cavity of the device is formed by a single self-coupled wire waveguide, random length fabrication errors in different parts (i.e., $SLR_1$ in Fig. 5(a), $SLR_2$ in Fig. 5(b), and feedback loop in Fig. 5(c),) will not induce any asymmetry in the filter shape. This yields a higher fabrication tolerance as compared with the coupled SLRs in Refs. [9, 10], which is particularly attractive for optical interleavers that require a flat-top symmetric filter shape. From Figs. 4(b)–(c) and Fig. 5, it can be seen that the slight changes in the structural parameters induced by fabrication disorders have no major impact on device performance.

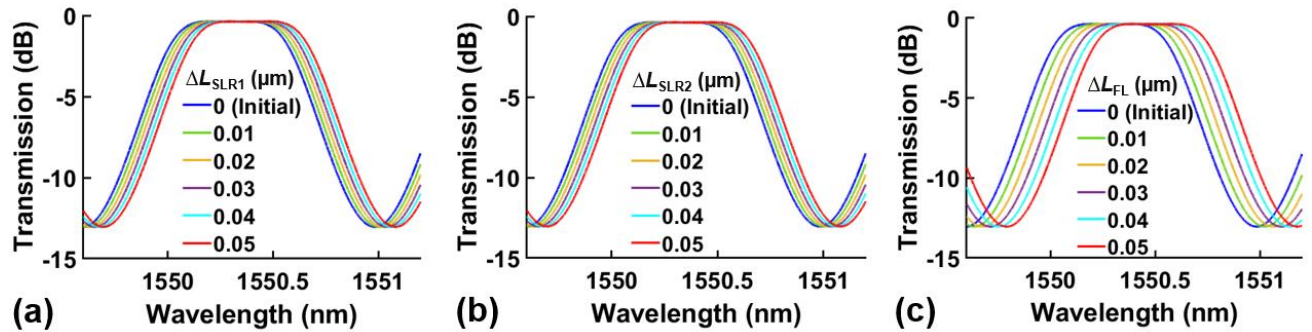

Figure 5. (a)–(c) Power transmission spectra for different $\Delta L_{SLRi}$ ($i$ = 1, 2) and $\Delta L_{FL}$, respectively. In (a)–(c), the structural parameters are kept the same as those in Fig. 4(a) except for the varied parameters.

## 5. CONCLUSION

We investigate theoretically photonic integrated filters based on two coupled SLRs with a feedback loop formed by a self-coupled optical waveguide. High performance filter functions including Fano-like resonances and wavelength de-interleaving are achieved by tailoring coherent mode interference in the device. Our design takes into account the device fabrication experience as well as the requirements for practical applications. The impact of device structural parameters on each filter function is analyzed to facilitate optimized performance. Versatile spectral responses, compact device footprint, and high fabrication tolerance make this approach highly promising for flexible response shaping in a wide variety of applications including potentially optical microcombs for advanced dispersion design for many applications. [25-97]

## REFRENCES